\documentclass{article}

\usepackage{arxiv}

\usepackage[utf8]{inputenc} 
\usepackage[T1]{fontenc}    
\usepackage{hyperref}       
\usepackage{url}            
\usepackage{booktabs}       
\usepackage{amsfonts}       
\usepackage{nicefrac}       
\usepackage{microtype}      
\usepackage{lipsum}
\usepackage{graphicx}

\title{Image-Like Graph Representations for Improved Molecular Property Prediction}

\author{
  Toni Sagayaraj \\
  Brown University \\
  antony\_sagayaraj@brown.edu \\
   \And
  Carsten Eickhoff \\
  Brown University \\
  carsten@brown.edu \\
}

\begin{document}
\maketitle

\begin{abstract}
Research into deep learning models for molecular property prediction has primarily focused on the development of better Graph Neural Network (GNN) architectures. Though new GNN variants continue to improve performance, their modifications share a common theme of alleviating problems intrinsic to their fundamental graph-to-graph nature. In this work, we examine these limitations and propose a new molecular representation that bypasses the need for GNNs entirely, dubbed CubeMol. Our fixed-dimensional stochastic representation, when paired with a transformer model, exceeds the performance of state-of-the-art GNN models and provides a path for scalability.
\end{abstract}


\section{Introduction}
The field of drug discovery has always been resource intensive due to the sheer number of candidates screened for any biological target. Even with modern methods of rational drug design and libraries of known bioactive compounds, the average development cost of a new drug hovers at around 2.8 billion dollars \cite{Wider2020}. In order to find better drug candidates, \textit{in-silico} molecular property prediction serves to vastly accelerate this process by improving the efficacy of screening. Due to their ability to leverage large datasets for improved performance, deep learning approaches have overtaken traditional methods like using molecular fingerprints or features from domain experts in conjunction with machine-learning algorithms \cite{mol_fingerprint}.

The application of deep neural networks to property prediction remains a nontrivial task due to the inherent graph structure of molecular data, which cannot be handled by traditional feedforward or convolutional architectures. Whereas standard neural networks operate on fixed-dimensional tensors of data for each input, graphs are almost always heterogeneous with varied numbers of nodes and edges. For this reason, the vast majority of advancements in property prediction have been achieved with Graph Neural Networks (GNNs), network architectures that restructure information propagation along the connectivity of graphs. These molecular graphs are represented by adjacency matrices along with feature vectors for each node and edge.

The majority of GNNs fall into two categories: Message Passing Neural Networks (MPNNs) and Graph Convolutional Networks (GCNs). Both architectures follow the same basic principles: Each layer of the network updates the representation of every node in the graph by aggregating information from its neighbors. At the end of the network, a readout function that compiles the node embeddings to produce a single output vector. Message Passing Networks differ slightly from GCNs by using message passing as their aggregation function, and include an extra update function before the next layer. Though these GNN variants have been specialized to work with molecular graphs, their methods of information propagation and readout functions hamper their expressive power and performance. We find that the primary bottlenecks of GNN performance arise from their graph-to-graph architecture, since constraining networks to transform sets of node embeddings poses challenges to learning deep representations that capture global structure and information.

In this paper, we propose CubeMol, a novel stochastic representation of molecular graphs that bypasses the restrictive nature of GNNs entirely. Our fixed-dimensional random walk representations not only allow for the application of pre-existing architectures like ConvNets and Transformers, but also implicitly encode multiple views of the molecule’s structure. Additionally, the stochastic nature of the representation provides natural data augmentation to improve model generalization. Our model exceeds the performance of state-of-the-art GNNs in multiple property prediction tasks.

\section{Limitations of Current GNNs}
\label{sec:headings}


\subsection{Long Distance Information Propagation}
Information exchange between two distant nodes has long been known to be difficult for GNNs to accomplish. In their recent work, Alon \& Yahav \cite{alon2021bottleneck} explain this phenomenon through architectural bottlenecks for message propagation, which leads to “over-squashing”. This stems from the fact that the receptive field of each node grows with each step of message aggregation. Given that communication between distant nodes requires many propagation steps, messages from exponentially more nodes are compressed into the same fixed-size vector. This limits the effective transmission range to the local neighborhood of any node, similar to the bottleneck for long-range interactions in RNN seq2seq models. Because global information is often vital for molecular properties, this aspect of GNNs limits their expressivity and performance for such tasks.

\subsection{Readout Functions And Expressivity}

Without the readout function, GNNs are graph-to-graph networks. Most property prediction models operate on sets of node embeddings, iteratively transforming them based on their connectivity with other nodes as well as the information contained in their edges. As molecular properties are often graph-based classification or regression tasks, generating a single fixed-dimensional representation from node embeddings is necessary. As a consequence, the diversity of information that can be captured from node embeddings is fundamentally limited. In addition to the standard average pooling function used, many pooling variants have been developed to mitigate this issue. Though set2set pooling mitigates this information loss via an LSTM, its sequential nature prevents the simultaneous consideration of all node embeddings \cite{vinyals2016order}. Given the known difficulty with communicating global information, further limiting the ability of nodes to assemble their local information could make a GNN more akin to boosting many shallow models rather than a single cohesive deep model. Given the bottlenecks associated with any pooling methods, a method of avoiding a many-to-one compression at all should further increase the expressive power of GNNs.

\subsection{Higher-Order and Multi-Scale Structures}

A network’s ability to capture molecular substructures and multi-scale features must govern its performance on global properties that rely on their interactions. For example, the proximity of different functional groups or their interactions with aromatic and conjugated systems greatly alters overall chemical properties. However, the other limitations discussed and the simple message aggregation schemes of most GNNs limit the ability to represent these features. If higher-order structures can only be captured by multiple message passing rounds, over-squashing directly limits the ability of GNNs to identify them. Furthermore, the fact that all molecular information is encoded in node representations leaves no trivial method of storing multi-scale features. If standard GNNs utilized these features for property prediction, they would need to be computed and incorporated into all node representations simultaneously. Otherwise, handcrafted architectural changes like graph coarsening are needed to account for hierarchical structures \cite{liao2019lanczosnet}.

\section{Related Work}

\paragraph{Improved Information Propagation}
The Directed MPNN (D-MPNN) was created with the goal of avoiding cyclical message passing between neighboring nodes \cite{yang2019dmpnn}. By operating on directed edge-based hidden representations and messages, D-MPNN also mitigates the over-squashing problem by limiting the redundant information exchange between neighbors. This is possible because a message $m_{vw}^{t+1}$ from node $v$ to $w$ does not depend on the reverse message $m_{wv}^t$ from the previous step, allowing bidirectional information flow without mixing the information like in vanilla MPNNs.

\paragraph{Leveraging Multi-Scale Information}
LanczosNet aims to better capture multi-scale information using a low-rank approximation of the graph Laplacian using the Lanczos algorithm \cite{liao2019lanczosnet}. From this representation, matrix powers can be easily calculated to gather multi-scale information. These features are then transformed by learnable spectral filters in order to increase the overall model capacity. Operation on a spectral representation of the graph also avoids the problems noted with the typical spatial aggregation methods at the cost of further computation.

\paragraph{Higher-Order Graph Structure}
The Path Augmented Graph Transformer (PAGTN) \cite{chen2019pathaugmented} and Path MPNN \cite{flamshepherd2020neural} both directly incorporate path information to capture higher-order graph properties. To accomplish this, both architectures expand the basic local aggregation methods of MPNNs to account for neighbors that are 2 or 3 nodes away, incorporating the path taken by concatenating bond features. Additional one-hot features such as whether atoms are in rings or part of functional groups further increases the information available to the network to better capture and represent global information.

\begin{figure}[h]
\includegraphics[width=0.35\textwidth]{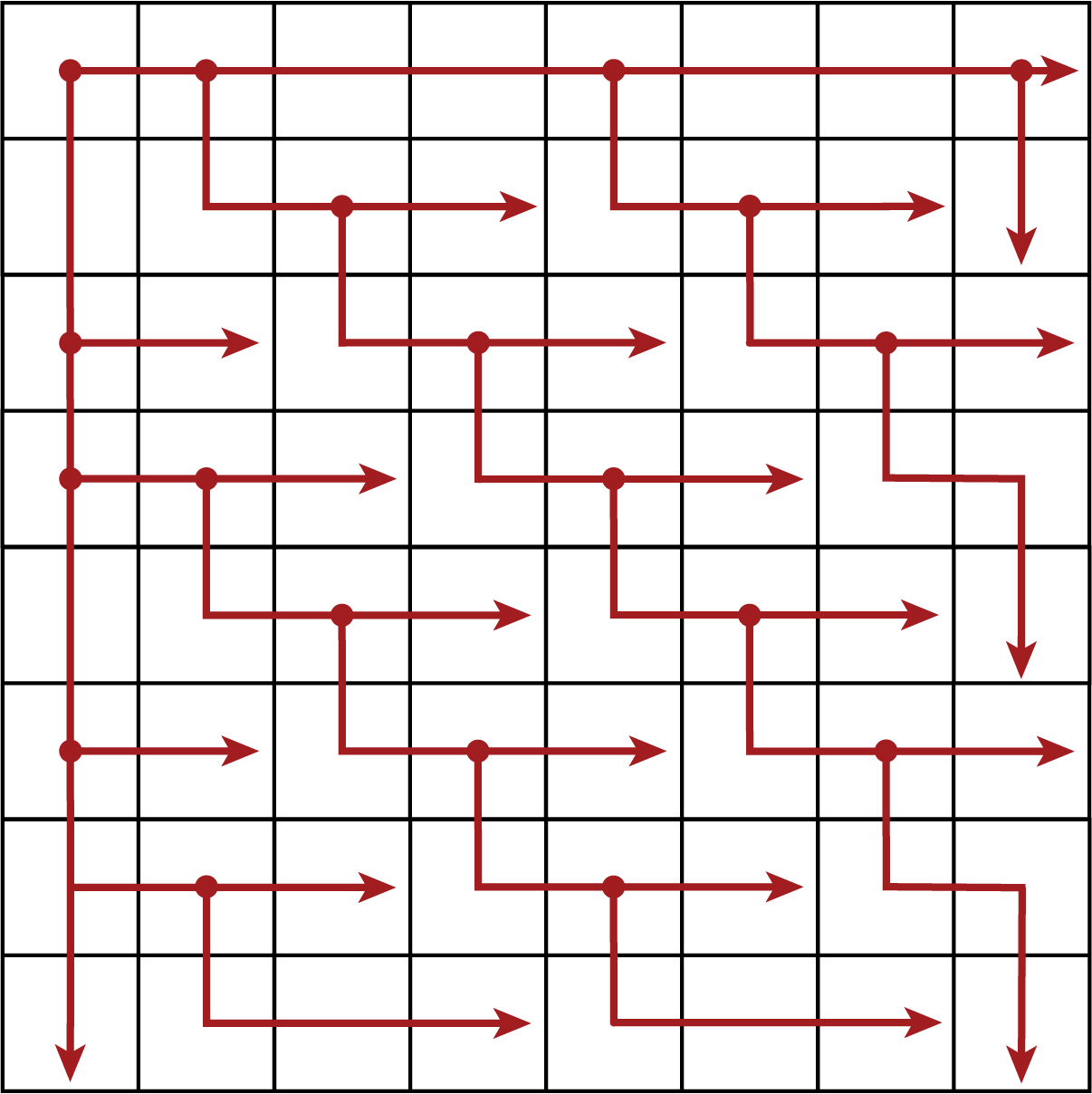}
\includegraphics[width=0.2\textwidth]{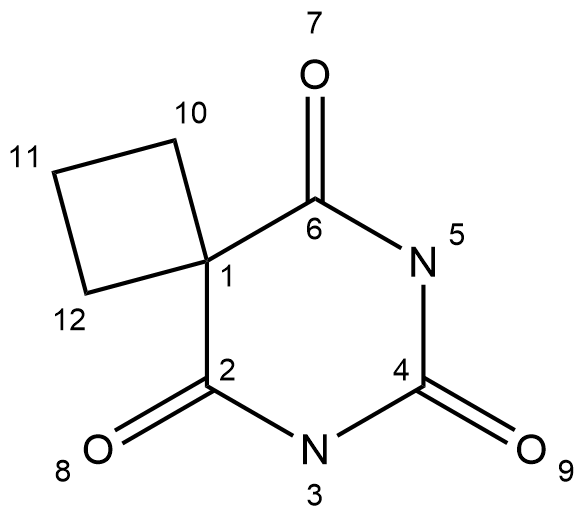}
\includegraphics[width=0.35\textwidth]{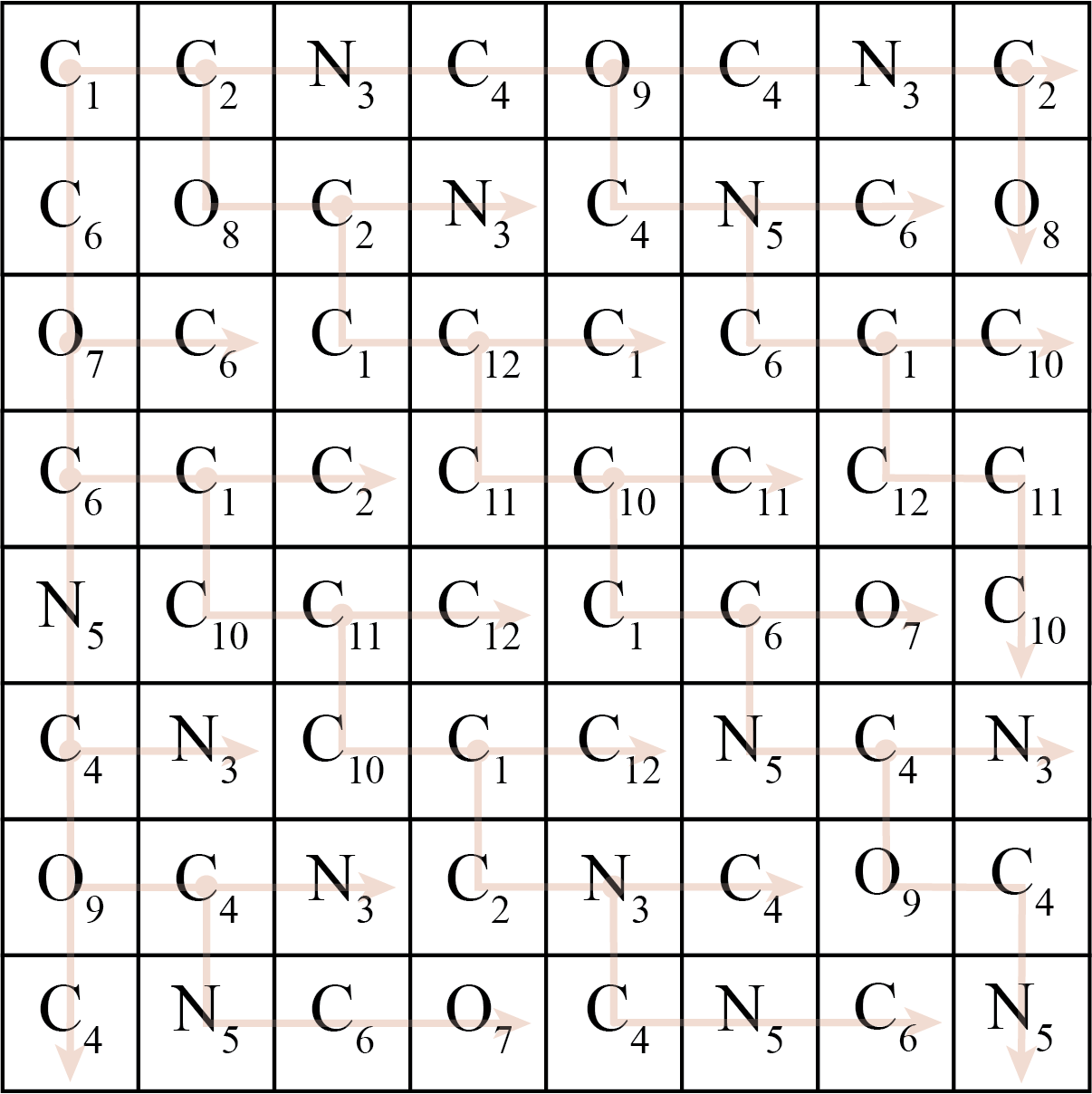} \\
(a) \hspace{4cm} (b) \hspace{4cm} (c)
\centering
\begin{centering}
\caption{Illustration of the CubeMol grid. a) An 8x8 grid demonstrating the branching random-walk pattern of CubeMol. The red nodes indicate where the random walks branch, and the arrows indicate the direction. b) Example molecule from ESOL. c) Sample grid of random walks on ESOL molecule.}
\end{centering}

\end{figure}

\section{Methods}

\subsection{CubeMol Representation}

Much like Cubist paintings seek to represent 3D forms by combining multiple perspectives, CubeMol combines many parallel random walks to produce a dense fixed-dimensional representation of a graph. Rather than a representation that explicitly indicates the connectivity of a graph, the random walks of CubeMol serve to provide redundant cues of both local and global structure. These walks branch out from one another and are arranged in a grid, forming an image.

After choosing the grid size, a random starting node is chosen for the top left corner. Then, branching random walks fill out the tiles in the pattern specified in Figure 1. Though there are many arbitrary choices for how the walks can fill the grid, we designed the pattern to capture multiple levels of locality. Because there are many branch points and locations where walks that have diverged remain close to each other, any given node in a tile is likely to have neighbors that are varied in distance on the original graph.

For each “pixel” representing a node in the graph, the channel dimensions contain both the node and edge features with the previous node in the walk. In molecular graphs, these correspond to the atom and bond features typically used by GNNs.

Given that the grid size is large enough for the random walks to traverse the entire graph, CubeMol should contain more than enough information to reconstruct the original graph. We avoid edge cases that could limit this information by preventing backtracking and branches that lead to the same neighbor if possible. The stochastic nature of this representation naturally provides data augmentation, as many distinct CubeMol tensors can be generated from a single graph.

\subsection{Vision Transformers for Graphs}

Recently, transformer models have proliferated the space of image classification thanks to the Vision Transformer (ViT) \cite{dosovitskiy2020image}. Dosovitskiy et al. found that pure transformers applied to sequences of image patches were able to achieve comparable results to state-of-the-art CNNs while requiring fewer computational resources. We chose to use a modified ViT model rather than a CNN because its self-attention is better suited to gather and compile information from the CubeMol representation.

Whereas the vanilla ViT segments the image into non-overlapping patches, we also use two types of additional patches shown in Figure 2. Edge patches cover information overlapping between two patches on the edges of the grid, and inner patches cover the intersections between four patches. These serve to facilitate learning the connection between these patches, since nodes on the edges between two patches have information that pertains to one another. We chose to add patches in this limited manner because of the $O(n^2d)$ computational complexity of attention on a sequence with n elements.

In addition to these patches, we maintained the standard of adding a learnable token to the sequence, whose final output would be used for the property prediction task. Before passing into the Transformer Encoder, all tokens were concatenated with a 1D learnable positional encoding.

\section{Experiments}

\subsection{Data}

All datasets were split into a 80/10/10 ratio for training, validation, and testing. To compare our model against baselines, we chose 

\begin{itemize}
\item QM8 \cite{QM8}: Dataset of 21786 organic molecules with 8 or fewer heavy atoms (C, O, N, F). Prediction of 12 electronic spectra values calculated by DFT.
\item ESOL \cite{ESOL}: Dataset of 1144 molecules for prediction of aqueous solubility
\end{itemize}
    
\subsection{Model Architecture \& Training}

For the CubeMol representation, we tested both 16x16 as well as 32x32 grids to check the impact of grid size on performance. For data augmentation, 100 sets of CubeMol representations were generated for each training dataset and grid size. On each epoch, a set is randomly chosen without replacement such that the model loops through all sets every 100 epochs.

For our transformer, we decided to use 6 attention layers, 16 attention heads, patch size of 4, 0.1 dropout on the transformer layers, and a final MLP layer with 1024 units. We tested the effectiveness of the additional patches, denoted by "+". All networks were trained for 1000 epochs, with a batch size of 64 and learning rate of $1\times 10^{-4}$ using the Adam optimizer. The validation set was evaluated every 10 epochs and the model checkpoint with the lowest validation loss was chosen for the final results on the test set.

\begin{figure}[t]
\includegraphics[width=0.33\textwidth]{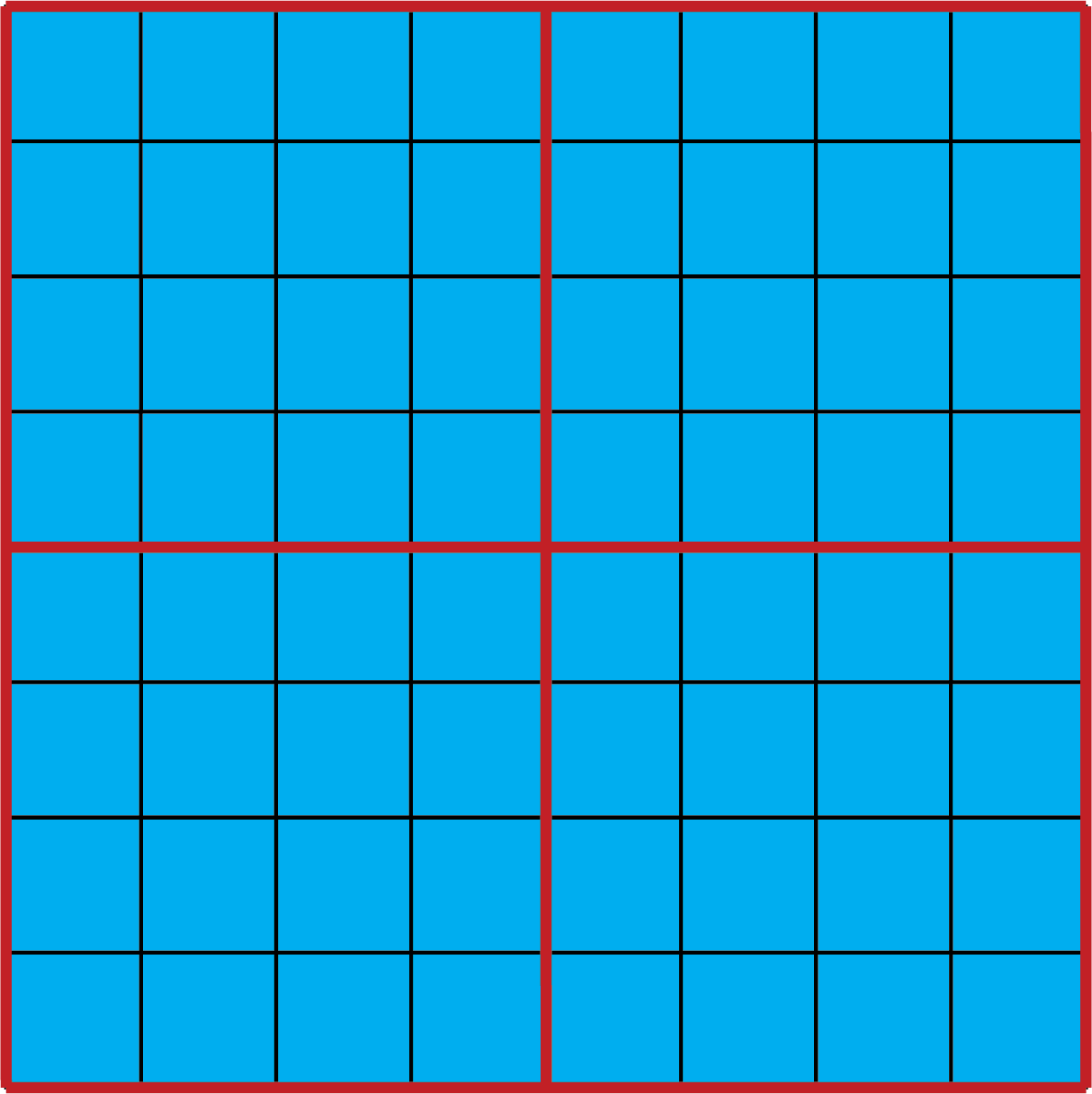}
\includegraphics[width=0.33\textwidth]{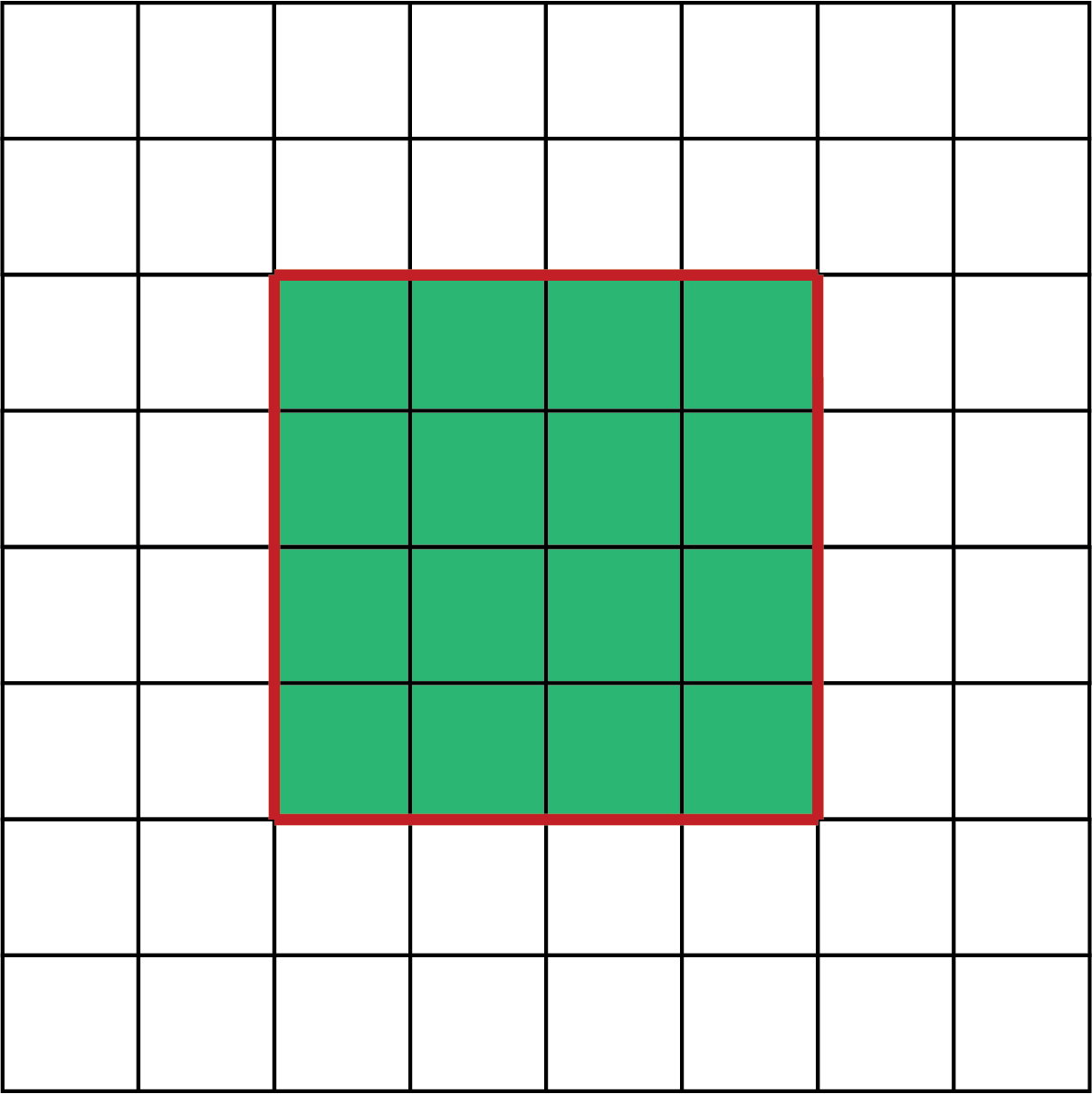}
\includegraphics[width=0.33\textwidth]{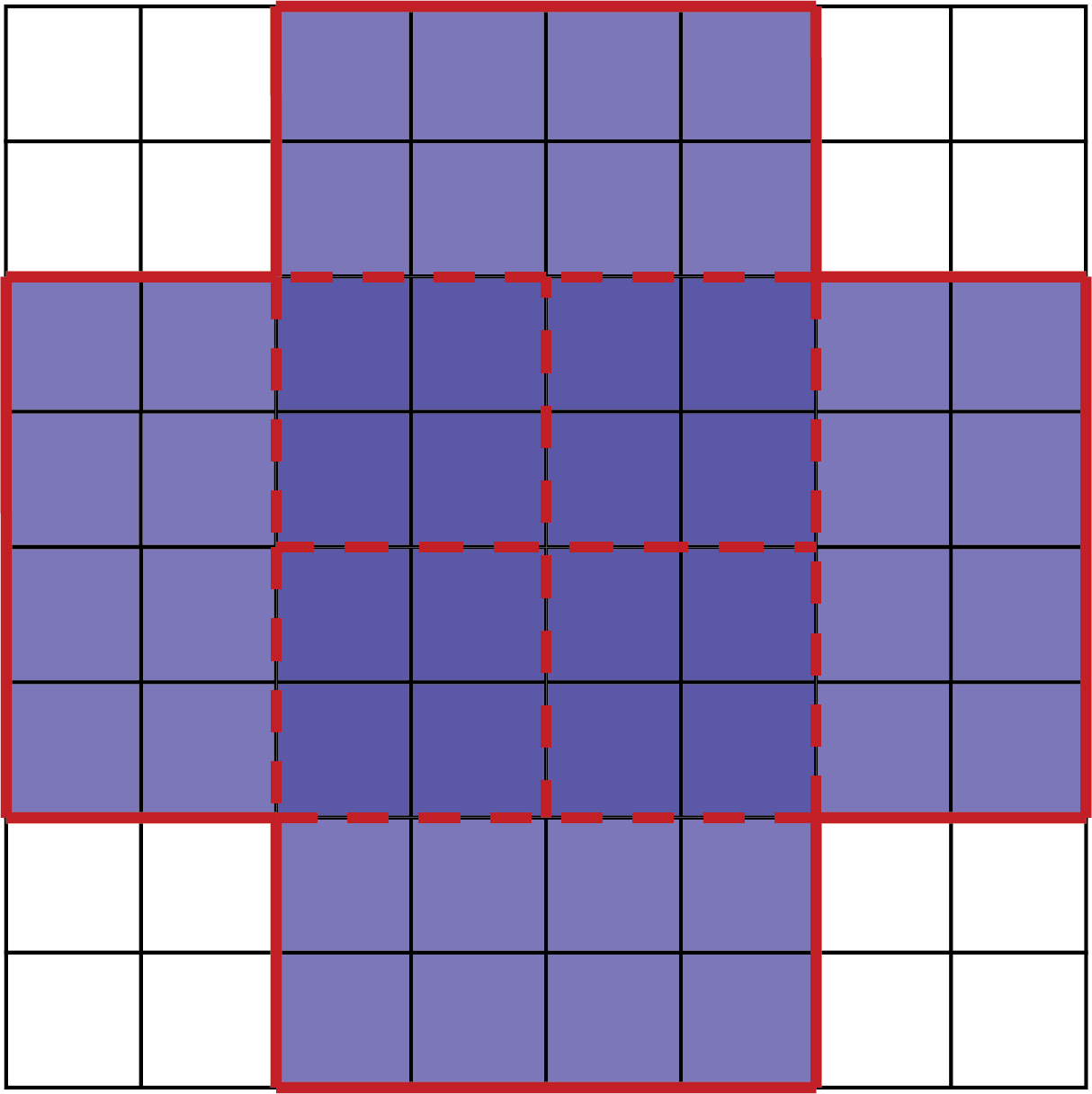} \\
(a) \hspace{5cm} (b) \hspace{5cm} (c)
\centering
\begin{centering}
\caption{Example of patches on 8x8 grid. a) Standard patches b) Inner overlap patches c) Edge overlap patches}
\end{centering}

\end{figure}


\section{Discussion}

The results of our experiments are shown in Table 1. On QM8, our model was able to achieve SOTA by a significant margin. This final performance, however, is likely not close to the true power of the CubeMol transformer. We did not perform any hyperparameter tuning, and model continued to improve even close to the end of the 1000 training epochs. This leads us to believe that larger, better-tuned models with appropriate learning rate schedules can lead to significant improvements. In fact, this behavior was somewhat implied in Dosovitskiy et al. \cite{dosovitskiy2020image} with their experiments. These mainly focused on pre-training the ViT on large amounts of data before fine-tuning on smaller image-recognition datasets. When directly trained on these smaller datasets like CIFAR-100, however, they reached nowhere near the same performance.

The performances of the baselines represented side-by-side reflect the limitations of GNNs stated prior. Of all networks, GGRNet \cite{shindo2019gated} performed the worst, ranking even below the MPNN \cite{gilmer2017neural} implemented by Flam-Shepherd et al. \cite{flamshepherd2020neural}. Though GGRNet implements a more complex MPNN, its readout function simply averages the node vectors and passes the result through an MLP for the result. The simpler MPNN uses the set2set readout instead, which is what likely lead to the better performance.

The mixed performance of D-MPNN \cite{yang2019dmpnn} on QM8 and ESOL may also be due to the basic sum readout function. However, the ability of D-MPNN to surpass the performance of the standard MPNN + set2set readout on QM8 indicates that global information propagation bottleneck is actually more impactful in this case. For the significantly larger molecules in ESOL, the sum readout greatly limits the expressivity of D-MPNN.

The Path MPNN \cite{flamshepherd2020neural} is arguably the best direct comparison to our CubeMol model. Though it does not use a transformer, the use of graph attention to aggregate messages across high order paths makes it analogous to the CubeMol transformer. The CubeMol Transformer can be viewed as reversing this construction, by explicitly laying out many paths across the molecule and then using a transformer to incorporate the information. Without the limitation of needing to store multi-scale information within the nodes of the graphs, the CubeMol likely surpasses the performance of the Path MPNN on QM8 due to a larger model capacity and expressive power.

Though the Path Augmented Graph Transformer (PAGTN) \cite{chen2019pathaugmented} uses a transformer to operate on graphs like an MPNN, its performance is once again limited by the readout function. Even with global attention to all nodes simultaneously, it appears that the sum operator used to aggregate hidden representations limits its power on both datasets compared to the Path MPNN, which uses the set2set model.

LanzcosNet \cite{liao2019lanczosnet} appears to be able to successfully utilize multi-scale information from the graph Laplacian and avoid limited model capacity, outperforming both the MPNN and PAGTN. However, the inaccuracies from the low-rank approximation provided by the Lanczos algorithm likely limits what the network can learn from the representation.

On ESOL, we believe our CubeMol models struggle due to the limited data, the large size of molecules, and the lack of path and structure features used in Path MPNN. Even still, the ability for the CubeMol models to scale to molecules with more than 50 heavy atoms indicates the room for optimization. For example, a tree decomposition method used by the Junction Tree Variational Autoencoder \cite{jin2019junction} would be able to greatly reduce the number of elements in the graph by generating a new vocabulary of nodes that includes cycles and functional groups.

\begin{figure}[t]
\begin{centering}
\begin{tabular}{ lccc } 
Dataset & QM8 & ESOL\\ 
Units & MAE in eV ($\times 10^{-3}$) & RMSE in log Mol/L \\ 
 \toprule
 GGRNet \cite{shindo2019gated}              & 13.94 $\pm$ 0.30 &  -- \\
 MPNN \cite{flamshepherd2020neural}  & 11.30 $\pm$ 0.31 & 0.47 $\pm$ 0.03 \\
 D-MPNN \cite{yang2019dmpnn}  & 11.00 $\pm$ 0.00 & 0.67 $\pm$ 0.05 \\
 PAGTN \cite{chen2019pathaugmented}    & 10.20 $\pm$ 0.30 & 0.55 $\pm$ 0.06  \\
 LanczosNet \cite{liao2019lanczosnet}              & 9.58 $\pm$ 0.14 &  --  \\
 Path MPNN \cite{flamshepherd2020neural}    & 8.70 $\pm$ 0.06 & \textbf{0.41 $\pm$ 0.02} \\
 \toprule
 CubeMol 16x16   & 9.35 $\pm$ 0.15 & 0.61 $\pm$ 0.02 \\
 CubeMol+ 16x16   & 8.76 $\pm$ 0.07 & 0.53 $\pm$ 0.05 \\
 \textbf{CubeMol+ 32x32} & \textbf{8.41 $\pm$ 0.05} & 0.45 $\pm$ 0.03 \\
 \toprule
\end{tabular}

\vspace{.25cm}
 Table 1 : Mean and std error on QM8, ESOL, and CEP for baselines and CubeMol models.
\vspace{.25cm}

\end{centering}
\end{figure}

\section{Conclusion and Future Works}

In this paper, we identified limitations to GNNs and proposed the CubeMol representation to avoid the use of GNNs entirely. The construction of CubeMol as fixed-dimensional representation composed of random walks directly tackles the issues faced by GNNs, which pertain to capturing global information and multi-scale structures. When combined with a patch-based transformer, the performance of CubeMol surpasses state-of-the-art GNNs on QM8.

The concepts introduced in this work can be further expanded to greatly improve the performance of deep learning models on graph data. Since the CubeMol representation is a tensor like any image, the entire suite of powerful image-recognition models can now be applied to molecular property prediction. These include all of the scaling benefits, improvements to the efficiency of self-attention, and architectures for deeper transformer models.

In hindsight, our particular CubeMol grid representation can be deconstructed to be much simpler and robust. Since the grid is eventually broken down into a sequence of patches for the transformer, these can be replaced by a sequence of random walks centered around each nodes. An algorithmic manner of effectively covering subgraphs with redundant copies around each node would both reduce the number of patches needed per graph and allow the transformer to dynamically scale to larger graphs.

\bibliographystyle{unsrt}  
\bibliography{references}  






\end{document}